
\input harvmac
\input psfig
%
%
%

\def\tilde{\widetilde}
\def\bar{\overline}
\def\hat{\widehat}
\def\*{\star}
\def\[{\left[}
\def\]{\right]}
\def\({\left(}		
\def\){\right)}

%
%

\def\frac#1#2{{#1 \over #2}}

\def\rvac{\hbox{$\vert 0\rangle$}}
\def\lvac{\hbox{$\langle 0 \vert $}}
\def\2pi{\hbox{$2\pi i$}}

\def\dsl{\raise.15ex\hbox{/}\kern-.57em\partial}
\def\Dsl{\,\raise.15ex\hbox{/}\mkern-.13.5mu D}
%
%
\def\th{\theta}

\def\ep{\epsilon}

%
%

%
\def\rvac{\hbox{$\vert 0\rangle$}}
\def\lvac{\hbox{$\langle 0 \vert $}}

\def\2pi{\hbox{$2\pi i$}}

\def\dsl{\raise.15ex\hbox{/}\kern-.57em\partial}
\def\Dsl{\,\raise.15ex\hbox{/}\mkern-.13.5mu D}
%
%
%
\font\numbers=cmss12
\font\upright=cmu10 scaled\magstep1
\def\stroke{\vrule height8pt width0.4pt depth-0.1pt}
\def\topfleck{\vrule height8pt width0.5pt depth-5.9pt}
\def\botfleck{\vrule height2pt width0.5pt depth0.1pt}
\def\Zmath{\vcenter{\hbox{\numbers\rlap{\rlap{Z}\kern
0.8pt\topfleck}\kern
2.2pt
                   \rlap Z\kern 6pt\botfleck\kern 1pt}}}
\def\Qmath{\vcenter{\hbox{\upright\rlap{\rlap{Q}\kern
                   3.8pt\stroke}\phantom{Q}}}}
\def\Nmath{\vcenter{\hbox{\upright\rlap{I}\kern 1.7pt N}}}
\def\Cmath{\vcenter{\hbox{\upright\rlap{\rlap{C}\kern
                   3.8pt\stroke}\phantom{C}}}}
\def\Rmath{\vcenter{\hbox{\upright\rlap{I}\kern 1.7pt R}}}
\def\Z{\ifmmode\Zmath\else$\Zmath$\fi}
\def\Q{\ifmmode\Qmath\else$\Qmath$\fi}
\def\N{\ifmmode\Nmath\else$\Nmath$\fi}
\def\C{\ifmmode\Cmath\else$\Cmath$\fi}
\def\R{\ifmmode\Rmath\else$\Rmath$\fi}

\Title{CLNS 95/1350; hep-th/9507053}
{\vbox{\centerline{Massless Boundary Sine-Gordon at the}
\centerline{Free Fermion Point:}
\centerline{Correlation and Partition Functions }
\centerline{with Applications to Quantum Wires}}}

\bigskip
\bigskip

\centerline{Robert M. Konik}
\medskip\centerline{Newman Laboratory}
\centerline{Cornell University}
\centerline{Ithaca, NY  14853}
\bigskip\bigskip

\vskip .3in

In this report we compute the boundary states (including the boundary
entropy) for the boundary sine-Gordon theory.  From the
boundary states, we derive both correlation and partition functions.
Through the partition function, we show that boundary sine-Gordon maps
onto a doubled boundary Ising model.  With the current-current correlators,
we calculate for finite system size
the ac-conductance of tunneling quantum wires with
dimensionless free conductance $1/2$ (or, alternatively interacting
quantum Hall edges at filling fraction $\nu = 1/2$).
In the dc limit, the results of \ref\kf{C. Kane and M. Fisher,
Phys. Rev. B46 (1992) 15233.} are reproduced.

\Date{7/95}

\noblackbox

\def\phib{\bar{\phi}}

\def\bh{\hat{\beta}}
\def\psib{\hbox{$\bar\psi$}}
\def\bs{\hbox{$\vert B\rangle$}}
\def\pz{\partial_z}
\def\pzb{\partial_{\bar z}}

\newsec{Introduction}

The massless sine-Gordon theory with an integrable boundary perturbation
provides a theoretical realization for a variety of statistical mechanical
systems.  Among those that have attracted the most attention are interacting
quantum Hall edges \ref\wen{X.G. Wen, Phys. Rev. B41 (1990) 12838.}
\ref\fls{P. Fendley, A. W. W. Ludwig and H. Saleur,  cond-mat/9408068.}
and tunneling in quantum wires \kf \ref\wa{E. Wong and I. Affleck,
Nucl. Phys. B 417 (1994) 403.}.
At present our understanding of the boundary sine-Gordon theory at
arbitrary coupling is limited
to knowledge of the boundary scattering matrices.  In
\ref\gz{S. Ghoshal and A. Zamolodchikov, Int. J. Mod. Phys. A9 (1994) 3841.}
these scattering matrices were derived from the imposition of the boundary
Yang-Baxter equation and the crossing-unitarity condition.
On one
occasion, knowledge of these matrices was sufficient for
the calculation of a physical parameter in these systems.
In \fls, by cleverly
coupling thermodynamic Bethe ansatz (TBA)
techniques to a Boltzmann transport equation, the conductance
of interacting quantum Hall edges at filling fraction $\nu =1/3$ was
computed.  However, in general, knowledge of correlation functions is needed
to access physical quantities.  But presently the general form of correlation
functions is unknown.

The computation of correlation functions is facilitated by knowledge of
boundary states \bs.  In two dimensional boundary field theory there are two
possible pictures in which to work: one with the boundary is in time, as
an initial condition, and one with boundary is in space.  It is in the
former picture that the boundary state
is used to calculate correlation functions.  Such
functions then have the general form:
\eqn\eIi{
\langle O_1 (x_1) \cdots O_n (x_n) \rangle = {\lvac O_1 (x_1) \cdots
O_n (x_n) \bs \over \langle 0 \bs}.}
As the boundary is in time, the Hilbert space of the theory remains unchanged
from its bulk counterpart.  As such, the boundary state \bs ~is expressible
in terms of these original states.

Integrability imposes powerful constraints on the form this expression
must take.  If $\{A_a(\th)\}_{a\in A}$ is a particle basis whose scattering
off the boundary is factorizable, and so is described by
\eqn\eIii{
A_a (\th ) = A_b(-\th ) R^b_a (\th ) \bs ,}
where $R^b_a(\th )$ is the boundary scattering matrix, the boundary state
takes the general form
\eqn\eIiii{
\bs = g \exp \left[ \int^{\infty}_0 d\th K^{ab}(-\th ) A_a(\th ) A_b(-\th )
\right ] \rvac ,}
where $K^{ab}(\th ) = R^b_{\bar a}(i\pi/2 - \th)$
and $A_{\bar a}$ denotes the charge conjugate of $A_a$.
However knowledge of \bs ~
does not guarantee the ability to write correlation
functions in a simple fashion.
The fields $O_i (x_i)$ of interest may well not
be simply expressible in the basis $\{A_a(\th )\}_{a\in A}$, that
is, the form
factors of $O_i (x_i)$ may well be non-zero for arbitrarily high particle
number.

This is the case for sine-Gordon theory.  The basis of
solitons $A_{\pm}(\th )$ scatters factorizably off the boundary.  However
the fields of interest, such as the current operator
or the Mandlestam fermions,
in general have mode expansions involving multi-soliton states.  Only at
the free fermion point is this not the case.  A solution to this problem
may be found in choosing a different diagonalization of the Hilbert space,
one that both induces factorizable scattering off the boundary and in
which the fields are simply expressed.  Such a basis may be found in
generalizations of the spinon fields found in
\ref\spin{P. Bouwknegt, A. Ludwig, and K. Schoutens, hep-th/9406020.}
\ref\spin1{D. Bernard, V. Pasquier, and D. Servan, hep-th/9404050.}
or in the anyon-super fermion
fields described by \ref\iso{S. Iso, hep-th/9411051.}.
However, at present it is not understood how to express sine-Gordon
in terms of such fields.
As such in the paper we focus on the free-fermion point
where these difficulties are absent.  Construction of the boundary states
and computation of the correlators in this limit will set the stage for
future calculations away from the free-fermion point.

At the free-fermion point, the motivating physical systems are both
interacting in the bulk: the quantum Hall edges are at filling fraction
$\nu = 1/2$ \foot{Though experiments have failed to observe
edge states at $\nu = 1/2$ \ref\wang{J. Wang and V. Goldman, Phys.
Rev. Lett. 67 (1991) 749.},  we point out the
application to quantum Hall edges to emphasize the underlying
theoretical unity
of quantum Hall edges with quantum wires.}
and the quantum wires having impurity-free conductance $e^2/2h$.
The free-fermion point describes interacting electrons because
these physical systems are not boundary problems but impurity problems, i.e.
the scattering point is in the bulk of the system.  To turn the impurity
problem into a boundary problem, the system is folded about the impurity.
In folding the system, the interacting electrons are transformed into
free ones.

The outline of this paper is as follows.  In section 2 we construct the
boundary states for the sine-Gordon theory on a cylinder.  This construction
comes in two parts.  We must both
compute the massless scattering matrices and the boundary entropy
$g = \lvac B \rangle$.  We do the latter in two ways, one using TBA
techniques, and one via a direct calculation of a limiting form
of the partition function.
Using these boundary states in section 3,
we compute relevant correlation functions and the partition
function in full generality.  This partition function is then
related to the partition function of a doubled boundary Ising model.
In section 4 we calculate the
expected ac-conductance of interacting quantum Hall edges/tunneling quantum
wires.  This calculation matches onto the dc-conductance calculated by
\kf.  In \kf ~the universal scaling form of the conductance was
calculated by
mapping the system onto a lattice model
which had been
solved previously by \ref\gui{F. Guinea, Phys. Rev. B 32 (1985) 7518.}.
The advantage of our calculation of the conductance lies both in that it
gives finite size corrections and that there is some chance it can
be generalized beyond the free-fermion point.

\newsec{Construction of Boundary States}

The action for the massive sine-Gordon with an integrable boundary
perturbation is
\eqn\eIIi{
S_{SG} = {1\over 8\pi} \int_R dx dt (\partial_z \Phi \partial_{\bar z} \Phi
+ 4\lambda \cos(\bh \phi) ) + {\alpha \over 4\pi} \int_B d\gamma
\cos ({\bh \over 2}(\Phi - \phi_o)) ,}
where $z = (t + ix)/2$, $\bar z = (t - ix)/2$, and $B$, the boundary, is
described via a parametric curve, $z = \gamma (y) , \bar z = \bar \gamma (y)$.
This curve circulates in a positive sense around the region $R$ over which
the bulk terms are integrated.  The mass term, $\lambda \cos(\bh \Phi )$, is
included (even though we are interested in the massless limit) to
mark out the basis of states we intend to employ.  Only solitons
scatter nicely
off the boundary, i.e. factorizably, and this is the basis the mass term
picks out.  The other alternative, a basis organized into conformal modules
labeled by
primary fields, leads to a form for \bs ~vastly more complicated than \eIiii.
Here $\alpha$ is a dimensionful parameter and $\phi_o$ a constant.  All
physical quantities are independent of the sign of $\alpha$.  Such a sign
change is implemented via $\phi_o \rightarrow \phi_o + 2\pi/\bh$.  But
the subsequent shift, $\Phi \rightarrow \Phi + 2\pi/\bh$, restores
the boundary
term leaving the bulk term invariant.

It is well understood that at $\bh = 1$ the bulk portion of the
sine-Gordon action is equivalent to a free Dirac fermion
\ref\rcole{S. Coleman, Phys. Rev. D11 (1975)
2088.}\ref\rmand{S. Mandelstam, Phys. Rev D11 3026 (1975).}.
Letting $\psi_\pm$ and $\psib_\pm$ be the left and right chiral
components of the Dirac fermion with U(1) charge $\pm 1$, the bosonisation
relations are
\eqn\eIIii{
\psi_\pm = \exp (\pm i \phi ) , ~~~~ \psib_\pm =
\exp (\mp i \bar\phi ) ,}
where $\phi$ and $\phib$ are the chiral components of the boson $\Phi$:
\eqn\eIIiii{\eqalign{
\phi (x,t)  &= {1 \over 2} \left(
\Phi (x,t)  + i \int^x_{-\infty} dx' \partial_t \Phi (x',t) \right); \cr
\bar\phi (x,t)  &= {1 \over 2} \left( \Phi (x,t)  - i \int^x_{-\infty} dx'
\partial_t \Phi (x',t) \right). \cr}}
The bulk Dirac action is then given by
\eqn\eIIiv{
S^{bulk}_D = {1\over 8\pi} \int_R dx dt \left( \psi_+ \pzb \psi_-
+ \psi_- \pzb \psi_+ + \psib_- \pz \psib_+ + \psib_+ \pz \psib_-
+ 2im (\psi_-\psib_+ - \psib_-\psi_+ ) \right).}
where a certain choice of gamma matrices\foot{
$\gamma^0 = \left( {0 \atop i} {-i \atop 0} \right)$,
$\gamma^1 = \left( {0 \atop -i} {-i \atop 0} \right)$}
has been used.

In \ref\akl{M. Ameduri, R. Konik, and A. LeClair, hep-th/9503088.}
the bosonisation
of the boundary term was developed.  This bosonization involves two
terms: one implementing the boundary conditions at the free point
($\alpha = 0$), and one implementing the conditions for the interpolating
perturbation ($\alpha \neq 0$).  At the free point, the fermions on the
boundary must satisfy
\eqn\eIIv{
(\partial_y \gamma)^{1/2} \psi_\pm = e^{\pm i\sigma}
(\partial_y \bar\gamma)^{1/2} \psib_\mp }
where  $\sigma$ is a constant.
The addition to the action that
implements this condition is
\eqn\eIIvi{
S^{free~bd.}_D = {i \over 8\pi} \int_B dy (e^{i\sigma}\psi_+\psib_+
+ e^{-i\sigma}\psi_-\psib_- ).}
To construct the interpolating action, \akl ~assumed
in the spirit of conformal
perturbation theory that the fields maintain the structure that they
possess at $\alpha = 0$.  With this assumption, the cosine perturbation
becomes
\eqn\eIIvii{\eqalign{
S^{int}_D = \alpha \int_B dy \cos ({\bh \over 2} (\Phi - \phi_o ) =
\alpha & \int_B dy \left[ e^{-i\phi_o/2}
((\partial_y \gamma)^{1/2} \psi_+ a_-
+ (\partial_y \bar\gamma)^{1/2} \psib_- a_+) + \right. \cr
& \left. e^{i\phi_0/2}((\partial_y \gamma)^{1/2} a_+ \psi_-
+ (\partial_y \bar\gamma)^{1/2} a_- \psib_+ ) \right],\cr}}
where $a_\pm$ are zero mode operators.  They equal
\eqn\eIIviii{
a_\pm = {1\over 4} \exp (\pm i (\pi/2 + \phi - \bar\phi ) ) .}
In the process of rewriting the boundary perturbation, $\alpha$ is
renormalized.  The full Dirac action with boundary is then
$S_D = S^{bulk}_D + S^{free~bd.}_D + S^{int.}_D$.

To derive the boundary state we place the boundary at $t = 0$.  Then
setting $\gamma (y) = iy$, $\bar\gamma (y) = -iy$, varying the full
action with respect to the fermions and zero modes, and then eliminating
the zero modes yields the following interpolating boundary conditions
at $t = 0$:
\eqn\eIIix{\eqalign{
& e^{i\phi_o}\psib_+ - i e^{i(\phi_o - \sigma)}\psi_- - i\psi_+
+ e^{-i\sigma}\psib_- = 0 \cr
& \partial_x (\psib_- - ie^{i\sigma}\psi_+) + \alpha^2 (i\psib_- +
\psi_-e^{i\phi_o} ) = 0 .\cr}}
Taking $\alpha = 0$ recovers the above free boundary condition.

To derive the boundary state, we interpret the above boundary condition to
vanish when acting upon it, i.e.
\eqn\eIIx{
(e^{i\phi_o}\psib_+ - i e^{i(\phi_o - \sigma)}\psi_- - i\psi_+
+ e^{-i\sigma}\psib_-)\bs = 0 ~~ {\rm etc.}}
\bs ~is expressed in terms of states from the fermionic Hilbert
space.  Thus we need to specify fermionic mode expansions.  It is here
we specialize to the massless case.  Massless mode expansions for fermions
on a cylinder of radius $2l$ are
\eqn\eIIxi{\eqalign{
\psi_\pm (z) & = \sqrt{1\over l} \sum^\infty_{n=1} \psi^\pm_n e^{-(n-1/2)z/l}
+ \psi^\pm_{-n} e^{(n-1/2)z/l}, \cr
\psib_\pm (\bar z) & =
\sqrt{1\over l} \sum^\infty_{n=1} \psib^\pm_n e^{-(n-1/2)\bar z/l}
+ \psib^\pm_{-n} e^{(n-1/2)\bar z/l}, \cr }}
where the modes satisfy the following algebra
\eqn\eIIxii{\eqalign{
& \{ \psi^\pm_n , \psi^\pm_m \} = \{ \psib^\pm_n , \psib^\pm_m \} =
\{ \psi_n , \psib_m \} = 0 \cr
& \{ \psi^-_n , \psi^+_m \}
= \{ \psib^-_n , \psib^+_m \} = \delta_{n+m,0} .\cr}}
The normalization of the mode expansions is fixed by insisting as
$z \rightarrow 0$
\eqn\eIIxiii{\eqalign{
\langle \psi_- (z) \psi_+ (0) \rangle &= {1 \over z} + \cdots ;\cr
\langle \psib_- (\bar z) \psib_+ (0) \rangle &= {1 \over \bar z} + \cdots .
\cr}}
The above mode expansions are anti-periodic, i.e. we are in the Neveu-Schwarz
sector. A priori we would expect a contribution to the boundary state
\bs ~ from the Ramond sector.  This is the case with the Ising model
(see \ref\chat{R. Chatterjee, hep-th/9412169.}
\ref\lma{A. LeClair, G. Mussardo, S. Skorik, and H. Saleur,
hep-th/9503227.}).  However as will be
shown, the boundary entropy for the Ramond sector is zero
in the massless limit,
and so this sector makes no contribution to \bs .

The boundary state takes the general form
\eqn\eIIxiv{
\bs ~ = g(\alpha , l) \exp \left[ \sum^\infty_{n=1}
a_n \psib^-_{-n}\psi^-_{-n} +
b_n \psib^+_{-n}\psi^+_{-n} +
c_n \psib^+_{-n}\psi^-_{-n} +
d_n \psib^-_{-n}\psi^+_{-n} \right] \rvac ,}
where $g(\alpha , l)$ is the boundary entropy.  In substituting the mode
expansions into \eIIx , expanding out \bs ~ to first order, and solving,
we find,
\eqn\eIIxv{\eqalign{
a_n &= ie^{-i\sigma}/(1+\lambda_n) ~~~~~ b_n = ie^{i\sigma}/(1+\lambda_n) \cr
c_n &= i\lambda_n e^{i\phi_o}/(1+\lambda_n) ~~~~
d_n = i\lambda_n e^{-i\phi_o}/(1+\lambda_n) , \cr}}
where $\lambda_n = \alpha^2 l /(n-1/2)$.  It now remains to compute
$g(\alpha , l)$.

The boundary entropy is easily derived from knowledge of the partition
function.  Consider the partition function for the theory on a cylinder
of radius $2l$ and length $R$.  (These dimensions for the cylinder
will remain the same throughout the paper.)
Interpreting the length R to be in the
time direction, the partition function in the limit $R \gg 2l$ is
given simply in terms of the g-factors:
\eqn\eIIxvi{
Z = g_a(\alpha , l) g_b(\alpha , l) e^{-RE_o} ,}
where $a$ and $b$ denote the
boundary conditions on the two ends of the cylinder and $E_o$ is
the ground state energy of the system.  The calculation of the partition
function in this limit will be done in two ways.  The first
calculates the partition function
directly while the second uses TBA techniques to access it.

To find the partition function, we keep the same limit $R \gg 2l$ but
reinterpret the axis of the cylinder as space.
The partition function then equals
\eqn\eIIxvii{
Z_\pm = {\rm Tr}(\pm1)^F e^{-4\pi l H}
= e^{-4\pi l E_o}\prod_k (1\pm e^{-4\pi lk}) ,}
where $\pm$ indicate anti-periodic (Neveu-Schwarz)/periodic (Ramond)
boundary conditions, $E_o = -1/2\sum_k 4\pi l k$,
and the product $\prod_k$ is over all allowed
modes.  To determine the allowed modes we apply the boundary conditions.

In this case there are two sets of boundary conditions: one at $t=0$ and
one at $t=R$:
\eqn\eIIxviii{\eqalign{
0 &= \psib_+ - i\psi_- -i\psi_+ + \psib_- \vert_{t=0} ;\cr
0 &= \partial_x (\psib_- - i\psi_+ ) +
\alpha^2 (i\psib_- + \psi_- )\vert_{t=0};\cr
0 &= \psib_+ + i\psi_- + i\psi_+ + \psib_- \vert_{t=R} ;\cr
0 &= \partial_x (\psib_- + i\psi_+ ) +
\alpha^2 (i\psib_- - \psi_- )\vert_{t=R}.\cr}}
The last two conditions arise because the boundary parametrization at
$t=R$ is $\gamma (y) = -iy$.  In these expressions both $\sigma$ and $\phi_o$
have been set to zero.  $\sigma$ can always be gauged away \gz\akl , and
$\phi_o \rightarrow \phi_o + c$ is a symmetry in the massless limit.

As it stands the fields $\psi_\pm$,$\psib_\pm$ are not independent.
A change of basis facilitates the determination of their interdependence.
We write
\eqn\eIIxix{
T_\pm = \psi_+ \pm \psi_- , ~~~~ \bar T_\pm = \psib_+ \pm \psib_- .}
The boundary conditions then become
\eqn\eIIxx{\eqalign{
0 &= \bar T_+ - i T_+  \vert_{t=0} ;\cr
0 &= \partial_x (\bar T_- + i T_- ) +
\alpha^2 (i\bar T_- + T_- )\vert_{t=0};\cr
0 &= \bar T_+ + i T_+  \vert_{t=R} ;\cr
0 &= \partial_x (\bar T_- - i T_- ) +
\alpha^2 (i\bar T_- - T_- )\vert_{t=R}.\cr}}
The boundary conditions clearly separate with this change of basis.

In the mode expansions
\eqn\eIIxxi{\eqalign{
T^\pm &= \sum_{k_\pm} t^\pm_{k_\pm} e^{ik_\pm (t + ix)} ,\cr
\bar T^\pm &= \sum_{k_\pm} {\bar t}^\pm_{k_\pm} e^{-ik_\pm (t - ix)} ,\cr}}
the above boundary conditions constrain the allowed values of $k_\pm$.
Substituting the mode expansions in, we find the following :
\eqn\eIIxxii{\eqalign{
0 &= 1 + X_\pm(k_\pm ) ;\cr
X_+ &= e^{-2iRk} ; \cr
X_- &= e^{-2iRk}{(\alpha^2 - ik)^2\over (\alpha^2 + ik)^2}. \cr}}
We see then that the $t_+$ modes are free while the
$t_-$ modes are interacting.

The partition function is then given by
\eqn\eIIxxiii{
\log Z_\pm = \sum_{k_+ > 0} \left[ 2\pi l k_+ +
\log (1\pm e^{-4\pi l k_+}) \right]  + (k_+ \rightarrow k_- ).}
These sums can be evaluated using Matsubara sum techniques developed in
\chat .
It is then straightforward to extract
expressions for the boundary entropy.
We relegate the details to an appendix.  The result is
\eqn\eIIxxiv{\eqalign{
\log g_\pm (\alpha , l ) &= {1 \over \pi} \int^\infty_0 {1 \over 1+ k^2}
\left[ \log (1\pm e^{-2\pi a k}) + \lim_{b\rightarrow 0}
\log (1\pm e^{-2\pi b k}) \right]  + {1 \mp 1 \over 4}\log (2) \cr
&= \cases{\log \left[ 2\sqrt{\pi} a^a /(\Gamma (a+1/2) e^a) \right] &$+$\cr
-\infty &$-$,\cr}}}
where again $a = 2 l \alpha^2$.  Thus in the Ramond sector, the boundary
entropy is zero.  These results can be compared with
calculations of the boundary entropy, $g^I$, of a boundary Ising model in
a magnetic field of strength $\alpha$.  In \lma\chat ~$g^I$ was found to be
\eqn\eIIxxv{\eqalign{
g^I_+(\alpha ) &= {\sqrt{2\pi} \over \Gamma (a+1/2)} \left( {a \over e}
\right)^a;\cr
g^I_-(\alpha ) &= 2^{1/4}{\sqrt{2\pi a} \over \Gamma (a+1)}
\left( {a \over e}\right)^a.\cr}}
Thus we have
\eqn\eIIxxvi{
g_\pm (\alpha ) = g^I_\pm (\alpha ) g^I_\pm (0).}
Hence we have verified at the level of boundary entropy the results
indicated in \gui: boundary sine-Gordon is equivalent
to two copies of a boundary Ising model, one copy in a boundary field of
strength $\alpha$ and one with zero field.  We will demonstrate this
equivalence at the level of partition functions in section 3.

We now go on to calculate the boundary entropy using TBA techniques.
To implement the TBA analysis, we need the boundary and bulk scattering
matrices to be diagonal.  This
is not the case for the sine-Gordon theory, but it is easily
rectified through a change of basis.  In the soliton/anti-soliton basis,
the boundary scattering is described by
\eqn\eIIxxvii{\eqalign{
A^{\dagger}_+ (\theta ) {\rm B} & = P^+(\th)A^{\dagger}_+ (-\th )
+ Q^+(\th )A^{\dagger }_-(-\th ) {\rm B} ; \cr
A^{\dagger}_- (\theta ) {\rm B} & = P^-(\th)A^{\dagger}_- (-\th )
+ Q^-(\th )A^{\dagger }_+(-\th ) {\rm B} ; \cr}}
where $A^\dagger_\pm$ are Faddeev-Zamolodchikov operators which create
solitons/anti-solitons.  In \akl ~the boundary scattering matrices
were found to be:
\eqn\eIIxxviii{\eqalign{
P^\pm (\th ) \equiv P (\th ) &=
\left( (1 - \gamma/2)\cosh (\th ) \right)/D(\th ); \cr
Q^\pm (\th ) \equiv Q (\th ) &= -i \sinh (2\th )/2 D(\th );\cr}}
where
\eqn\eIIxxix{\eqalign{
\gamma &= 2\alpha^2/m; \cr
D(\th ) &= i\gamma \cosh ({\th +i\pi/2\over 2}) \sinh ({\th - i\pi/2\over 2})
- \cosh^2 (\th ) , \cr }}
and again both $\sigma$ and $\phi_o$ have been set to zero.
Although we are interested in the massless limit, it is easier to work
in the massive case, taking $m \rightarrow 0$ only at the end.

To diagonalize the scattering off the boundary, we introduce the operators
in analogy with \eIIxix :
\eqn\eIIxxx{
T^\dagger_{\pm} = A^\dagger_+ \pm A^\dagger_- .}
As is easily seen, this basis scatters diagonally off the boundary:
\eqn\eIIxxxi{
T^\dagger_{\pm}(\th )\bs = \left(
P(\th ) \pm Q(\th ) \right) T^\dagger_{\pm}(-\th )\bs
\equiv R_\pm (\th ) T^\dagger_\pm (-\th )\bs .}
It is also easily seen that scattering in this new basis is unitary, i.e.
$R_\pm(\th )R_\pm (-\th ) = 1$, important as the TBA
analysis requires it.

\def\rhot{\tilde\rho}
We again keep the same limit $R \gg 2l$ and the interpretation of
the axis of the
cylinder as space.  Following
\ref\zambo{Al. B. Zamolodchikov, Nucl. Phys. B342 (1990) 695.},
the system can be described as a widely spaced
set of n particles of definite rapidities, $\theta_i$, types $c_i$, and
located in regions $x_i$.  Because the particles are widely spaced,
the particles move as free ones and we can ignore off-mass shell effects.
Hence we can describe the system via a wavefunction
$\Psi (\theta_i,c_i, x_i)$.
Knowledge of the scattering matrices allows to constrain the
wavefunction arrived at through the interchange of two adjacent
particles:
\eqn\eIIxxxii{\eqalign{
&\Psi (\cdots;\theta_i,c_i,x_i;\theta_{i+1},c_{i+1},x_{i+1};\cdots ) \cr
&~~~~~~~~~~~~~~~~~~~~~~~~= S_{i,i+1} (\theta_i - \theta_{i+1})
\Psi (\cdots;\theta_{i+1},c_{i+1},x_i;\theta_i,c_i,x_{i+1};\cdots ) }}
or through the scattering of a particle off the two boundaries a and b:
\eqn\eIIxxxiii{\eqalign{
\Psi (x_1,\theta_1,c_1,\cdots ) &=
R^a_1(\theta_1)\Psi (x_1,-\theta_1,c_1,\cdots );\cr
\Psi (\cdots,x_n,\theta_n,c_n) &= (R^b_n)^{-1}(\theta_n)
\Psi (\cdots,x_n,-\theta_n,c_n),\cr}}
where $S_{ij}$ describes scattering of particles of types $c_i$ with
$c_j$ (only two indices are needed as the scattering is diagonal) and
$R_i$ describes scattering of particles of type $c_i$ off the boundary.
Scattering the i-th particle of rapidity $\theta_i \neq 0$ up the cylinder,
back down, and then up to its original location leads to the quantization
condition:
\eqn\eIIxxxiv{
e^{-2im_iR\sinh \theta_i}\prod_{i\neq j} S_{ij}(\theta_i-\theta_j)
S_{ij}(\theta_i+\theta_j)R^a_{i}(\theta_i)
R^b_{i}(\theta_i) = 1.}
Writing $\rho_i(\theta)$ as the density of levels for particles of type
$c_i$, and $\rhot_i (\theta)$ as the density of occupied states for particles
of type i, this quantization condition can be recast as
\eqn\eIIxxxv{\eqalign{
2\pi \rho_i (\theta ) &= mR\cosh (\theta ) + i\sum_j\int^\infty_{-\infty}
d\theta ' \rhot (\theta ') \partial_\theta\log S_{ij}(\theta - \theta ') \cr
& + {i\over 2} \partial_\theta \log (R^a_i(\theta ) R^b_i(\theta ) )
- {i \over 2} \partial_\theta \log S_{ii}(2\theta ) +
\pi\delta (\theta ).\cr}}
$\theta = 0 $ is not an allowed solution to the quantization
condition.  Thus $\delta (\theta )$ is included (by hand)
to remove this contribution.

\def\rg{\rho_i}
\def\rgt{\tilde\rho_i}

If the system is bosonic, its entropy is given by
\eqn\eIIxxxvi{
S_b = \sum_{i} \int^\infty_{-\infty} d\theta \left[
(\rgt + \rg )\log (\rg + \rgt ) - \rg\log\rg - \rgt\log\rgt \right],}
while if it is fermionic, the entropy is
\eqn\eIIxxxvi{
S_f = \sum_{i} \int^\infty_{-\infty} d\theta \left[
(\rgt - \rg )\log (\rg - \rgt ) + \rg\log\rg - \rgt\log\rgt \right].}
In either case the system's energy is
\eqn\eIIxxxviii{
H = \sum_{i} \int^{\infty}_{-\infty} d\theta m_i R \cosh (\theta )
\rgt .}
Finding the extremum of the free energy
\eqn\eIIxxxix{
\log Z_{b/f} = -4\pi l F_{b/f} = -4\pi l H + S_{b/f} ,}
by varying $\rgt$ and $\rg$ and using the quantization condition leads
to the following set of integral equations:
\eqn\eIIil{\eqalign{
\ep^{b/f}_i(\theta ) &= 4\pi l m_i \cosh \theta \mp {1 \over 2\pi i}
\sum_{j} \int^\infty_{-\infty} d\theta ' \partial_{\theta '}
\log S_{ij}(\theta ' -\theta)
\log (1 \mp e^{-\ep^{b/f}_j (\theta ')}) ;\cr
\log Z_{b/f} &= \mp \sum_i \int^\infty_{-\infty} {d\theta \over 2\pi}
\log (1 \mp e^{-\ep^{b/f}_i(\theta )}) ( mR \cosh \theta  + \cr
&~~~~~~~~~~~~~~~~~~~~~~~~~{i\over 2}\partial_\theta
\log (R^a_i(\theta ) R^b_i(\theta ))
- {i \over 2}\partial_\theta \log S_{ij}(2\theta ) +
\pi\delta (\theta )).}}
where we have introduced pseudo-energies, $\ep^{b/f}_i$, given by
\eqn\eIIili{
\ep^{b/f}_i = {\rgt \over \rg \pm \rgt }.}
The pieces in the equation for $\log Z_{b/f}$ corresponding to the
boundary entropy are those that do not scale with R.  Thus
\eqn\eIIilii{\eqalign{
\log g^{b/f}_a + \log g^{b/f}_b = & \mp \sum_i \int^\infty_{-\infty}
{d\theta \over 4\pi} \log (1 \mp e^{-\ep^{b/f}_i(\theta )}) \times \cr
&(i\partial_\theta \log (R^a_i(\theta ) R^b_i(\theta ))
- i\partial_\theta \log S_{ii}(2\theta ) +
2\pi\delta (\theta )).\cr}}

In the Neveu-Schwarz sector of the theory, the fermions fill the levels
as fermions.  Thus $\log g_{NS} = \log g_f$.  Examining the partition
function in \eIIxvii ~for the Ramond sector, we see it is equivalent to
the inverse of a bosonic partition function.  So $\log g_R = -\log g_b$.
As the theory is trivial in the bulk (i.e. $S = -1$), $\ep^{b/f}_i
(\theta ) = 4\pi l m \cosh \theta $.  So the
boundary entropy for the two sectors is given by
\eqn\eIIiliii{\eqalign{
\log g_\pm \equiv \log g_{NS/R} = \int^{\infty}_{-\infty} d\theta
&\log (1 \pm
e^{-4\pi l m\cosh \theta}) \times \cr
&\left( {i \over 4\pi}\partial_\theta
\log (P^2 (\theta ) - Q^2 (\theta ) ) + {1\over 2}\delta
(\theta )\right).\cr}}
This expression differs by an overall sign from that used in \lma
{}~to compute the boundary entropy of the boundary Ising model.
This change in sign results
from a difference in sign conventions used for the mass in deriving
the scattering matrices.
Doing the integrals and taking the massless limit leads to
\eqn\eIIiliv{\eqalign{
\log g_+ &= \log \left[ 2\sqrt{\pi}a^a / \Gamma (a+1/2) e^a \right] ; \cr
\log g_- &= -\infty , \cr}}
where $a = 2 l\alpha^2 $.  We thus see that the TBA analysis reproduces
exactly
the results of the direct calculation.  In general, TBA only can reproduce
the boundary entropy up to a constant.  Corrections arise both from
the use of Stirling's formula in the expressions for the entropy and
from off-mass shell effects.
But apparently, TBA is exact in this case.

\newsec{Correlation Functions and Partition Functions}

\subsec{Correlation Functions}

\def\xi{(x,\tau )}
\def\xj{(x', \tau ' )}
\def\dt{\tau - \tau '}
\def\mdt{\tau ' - \tau}
Given below are the two-point functions together with the current-current
correlators.  First we give the two-point functions unaffected by the
boundary,
the left-left and right-right fermionic correlators:
\eqn\eIIIi{\eqalign{
\langle \psib^\pm\xi \psib^\mp\xj \rangle & = {1 \over l}
\left[\theta (\dt ) {e^{-\bar s /4l} \over 1 - e^{-\bar s/2l}}
- \theta (\mdt ) {e^{\bar s /4l} \over 1 - e^{\bar s/2l}} \right] ;\cr
\langle \psi^\pm\xi \psi^\mp\xj \rangle & = {1 \over 2l}
\left[\theta (\dt ) {e^{-s /4l} \over 1 - e^{-s/2l}}
- \theta (\mdt ) {e^{s /4l} \over 1 - e^{s/2l}} \right] ,\cr }}
where $s = \tau - \tau ' + i(x-x')$.
The right-left left-right fermionic two point
functions, on the other hand, are affected by the boundary.  No longer zero,
they couple to the two particle contribution \bs , and are given by:
\eqn\eIIIii{\eqalign{
\langle \psib^\pm\xi \psi^\pm\xj \rangle & =
-{1 \over l} \sum_{n\in Z^+} {ie^{\mp i \sigma} \over 1 + \lambda_n}
e^{-(n-1/2)\bar y /2l} ;\cr
\langle \psi^\pm\xi \psib^\pm\xj \rangle & =
{1 \over l} \sum_{n\in Z^+} {ie^{\mp i \sigma} \over 1 + \lambda_n}
e^{-(n-1/2)y /2l};\cr
\langle \psib^\mp\xi \psi^\pm\xj \rangle & =
-{1 \over l} \sum_{n\in Z^+} {ie^{\pm i \phi_o} \lambda_n
\over 1 + \lambda_n} e^{-(n-1/2)\bar y /2l} ;\cr
\langle \psi^\pm\xi \psib^\mp\xj \rangle & =
{1 \over l} \sum_{n\in Z^+} {ie^{\pm i \phi_o} \lambda_n \over 1 + \lambda_n}
e^{-(n-1/2) y /2l} .\cr}}
where $y= \tau + \tau ' + i(x - x')$.  The current-current correlators are
then given in terms of these two-point functions:

\eqn\eIIIiii{\eqalign{
\langle j_r \xi j_r \xj \rangle & = {1 \over (4\pi )^2}
\langle \psib^- \xi \psib^+ \xj \rangle
\langle \psib^+\xi \psib^- \xj \rangle \cr
& = {1 \over (4l\pi )^2}
\left[\theta (\dt ) {e^{-\bar s /2l} \over (1 - e^{-\bar s/2l})^2}
+ \theta (\mdt ) {e^{\bar s /2l} \over (1 - e^{\bar s/2l})^2} \right] ;\cr
\langle j_r \xi j_r \xj \rangle & = {1 \over (4\pi )^2}
\langle \psib^- \xi \psib^+ \xj \rangle
\langle \psib^+\xi \psib^- \xj \rangle \cr
& = {1 \over (4l\pi )^2}
\left[\theta (\dt ) {e^{- s /2l} \over (1 - e^{-s/2l})^2}
+ \theta (\mdt ) {e^{s /2l} \over (1 - e^{ s/2l})^2} \right] ;\cr}}

\eqn\eIIIiv{\eqalign{
\langle j_r \xi j_l \xj \rangle & = {1 \over (4\pi )^2}
\left( \langle \psib^+ \xi \psi^- \xj \rangle
\langle \psib^-\xi \psi^+ \xj \rangle \right. \cr
& ~~~~~~~~~~~~~~~~~ \left. - \langle \psib^+ \xi \psi^+ \xj \rangle
\langle \psib^-\xi \psi^- \xj \rangle \right) \cr
& = {1 \over (4l\pi)^2}\sum_{k,k'\in Z^+} {1 - \lambda_k\lambda_{k'} \over
(1 + \lambda_k)(1 + \lambda_k')} e^{-(k+k'-1)\bar y/2l};\cr
\langle j_l \xi j_r \xj \rangle & = {1 \over (4\pi )^2}
\left( \langle \psi^- \xi \psib^+ \xj \rangle
\langle \psi^+\xi \psib^- \xj \rangle \right. \cr
& ~~~~~~~~~~~~~~~~~ \left. - \langle \psi^- \xi \psib^- \xj \rangle
\langle \psi^+\xi \psib^+ \xj \rangle \right) \cr
& = {1 \over (4l\pi)^2}\sum_{k,k'\in Z^+} {1 - \lambda_k\lambda_{k'} \over
(1 + \lambda_k)(1 + \lambda_k')} e^{-(k+k'-1) y/2l};\cr }}

\subsec{Calculation of Partition Functions}

\def\Z{Z_{\alpha '\alpha}}
We have already demonstrated that at the level of boundary entropy, the
boundary sine-Gordon is equivalent to two copies of boundary Ising, one
copy with a free boundary and one copy with a boundary in a magnetic field
of strength $\alpha$.  We now show this equivalence holds at the level
of partition functions.

With time along the axis of the cylinder, the partition function in the
presence of two boundaries is given by
\eqn\eIIIv{
\Z = \langle {\rm B} (\alpha ' ) \vert
e^{-H R} \vert {\rm B} (\alpha ) \rangle , }
where $H$ is the Hamiltonian on the cylinder:
\eqn\eIIIvi{\eqalign{
H &= {1 \over 2l} (L_0 + \bar L_0 - {c \over 12} ) \cr
&= \sum_{n\in Z^+}{ n - 1/2 \over 2l}
\left( \psi^-_{-n}\psi^-_{n} + \psi^+_{-n}\psi^+_{n} + \psib^-_{-n}
\psib^-_{n}
+ \psib^+_{-n}\psib^+_{n} \right) - {1 \over 24 l}\cr.}}
To compute the inner product in \eIIIiv, we use the formula
\eqn\eIIIvii{
\lvac e^{(\tilde a , Ma)}e^{(a^\dagger , N \tilde a^\dagger )} \rvac =
\det (1 + NM) ,}
where $(\tilde a , Ma) = \sum_{nm} \tilde a_n M_{nm} a_m$,
$\{ a_n , a^\dagger_m \} = \{ \tilde a_n , \tilde a^\dagger_m \} =
\delta_{n+m,0}$, and $ \{ \tilde a_n , a^\dagger_m \} = 0$.
We thus obtain
\eqn\eIIIviii{
\Z = g_+(\alpha ) g_+(\alpha ') q^{-1/24} \prod^\infty_{n=1} (1 + q^{n-1/2})
(1 + {(1-\lambda_n)(1-\lambda_n ') \over (1 + \lambda_n)(1+\lambda_n ')}
q^{n-1/2} ), }
where $q = e^{-R/l}$ and $\lambda_n = \alpha^2 l / (n-1/2)$.

The partition function for an Ising model with magnetic fields $\alpha$,
$\alpha '$ on the boundaries was computed in \lma:
\eqn\eIIIix{\eqalign{
Z^I_{\alpha \alpha '} =& {1 \over 2} q^{-1/48} g^I_+ (\alpha)
g^I_+(\alpha ') \prod^\infty_{n=1} \left( 1 + a_{n-1/2}(\alpha )
a_{n-1/2}(\alpha ') q^{n-1/2} \right)\cr
&+ {1 \over 2} {\rm sgn} (\alpha \alpha ' ) q^{1/24} g^I_- (\alpha)
g^I_-(\alpha ') \prod^\infty_{n=1} \left( 1 + a_{n-1}(\alpha )
a_{n-1}(\alpha ') q^{n-1} \right) ,}}
where
\eqn\eIIIx{\eqalign{
a_n(\alpha ) & = {1 - \alpha^2 l /n \over 1 + \alpha^2 l /n} ,\cr
a &= 2 l \alpha^2 ,\cr}}
and $g^I_\pm$ are given in \eIIxxiii.
The first term in $Z^I_{\alpha \alpha '}$ is the contribution from the
Neveu-Schwarz sector of the theory and the second term is the contribution
from the Ramond sector (zero in the case of boundary sine-Gordon).
We can then make the following identification:
\eqn\eIIIxi{
\Z = 2 Z^I_{00} \left( Z^I_{\alpha\alpha '} + Z^I_{\alpha, -\alpha'}
\right),}
where we have used the fact $g_+(\alpha ) = g^I_+(\alpha ) g^I_+(\alpha ')$.
Summing $Z^I_{\alpha \alpha '}$ with $Z^I_{\alpha,- \alpha '}$ mods out
the Ramond sector of the Ising theory.  This cancellation reflects the
indifference of boundary sine-Gordon to the sign of $\alpha $.

At the conformal boundary points (i.e. $\alpha = 0 $ and $\alpha =\infty$)
the partition function of the boundary sine-Gordon can be
expressed in terms of the $c=1/2$ characters.  The corresponding
expressions for  boundary
Ising are well known \ref\car{J. Cardy, Nucl. Phys. B 324 (1989) 581.}:
\eqn\eIIIxii{
Z^I_{ab} = \sum_{ij} n^i_{ab} S^j_i X_j (q) ,}
where $a$ and $b$ label the boundary conditions, $S$ governs modular
transformation of the $c=1/2$ characters $X_j$, $j= 0,1/2,1/16$, and
$n^i_{ab}$ is the number of times that the irreducible representation of
highest weight i appears in the spectrum of the cross-channel Hamiltonian
with boundary conditions $a$ and $b$.  S is given by
\eqn\eIIIxiii{
S = {1\over 2} \pmatrix{1&1&-\sqrt{2}\cr
1&1&\sqrt{2}\cr
\sqrt{2}&-\sqrt{2}&0\cr},}
and the non-zero values of $n^i_{ab}$ (all equal to 1) are
$n^0_{\pm \infty,\pm\infty},n^0_{00},n^{1/2}_{00},n^{1/2}_{\pm\infty,
\mp\infty}$, and $n^{1/16}_{\pm\infty,0}$.  The $c=1/2$ characters are
given explicitly by
\eqn\eIIIxiv{\eqalign{
\chi_0(q) &= { q^{-1/48} \over 2} \left[ \prod^\infty_{n=1} (1+q^{n-1/2})
+ \prod^\infty_{n=1} (1-q^{n-1/2}) \right] ;\cr
\chi_{1/2}(q) &= { q^{-1/48} \over 2} \left[ \prod^\infty_{n=1} (1+q^{n-1/2})
- \prod^\infty_{n=1} (1-q^{n-1/2}) \right] ;\cr
\chi_{1/16}(q) &= q^{1/2} \prod^\infty_{n=0} (1+q^n) .\cr}}
Then using \eIIIxi ~and \eIIIxii , the conformal
points of the partition function for boundary sine-Gordon
equal:
\eqn\eIIIxv{\eqalign{
Z_{00} &= 4\left( \chi_0(q) + \chi_{1/2}(q) \right)^2 ;\cr
Z_{0,\pm\infty} &= 2^{3/2}\left(\chi_0^2(q) - \chi_{1/2}^2(q) \right) ;\cr
Z_{\pm\infty,\pm\infty} &= Z_{\pm\infty,\mp\infty} =
2\left( \chi_0(q) + \chi_{1/2}(q) \right)^2 .\cr}}
These formulas can be directly verified by taking the appropriate limits
in \eIIIviii .

\newsec{Calculation of AC-Conductance}
\def\pl{\hbox{$\phi_L$}}
\def\pr{\hbox{$\phi_R$}}
\def\pe{\hbox{$\phi_e$}}
\def\po{\hbox{$\phi_o$}}
\def\plr{\hbox{$\phi_r$}}
\def\pll{\hbox{$\phi_l$}}

Using the results of the previous sections, we now go on to calculate the
conductance of a tunneling quantum wire with free conductance $e^2/2h$,
or equivalently, two interacting quantum Hall edges at $\nu = 1/2$.  We
begin by mapping these systems onto the boundary sine-Gordon at
$\bh = 1$.  The Hamiltonian
for an impurity free wire/two non-interacting edges is
\eqn\eIVi{
H_0 = -{v \over 4\pi\nu} \int^R_{-R} dx (\partial_x \phi_L)^2 +
(\partial_x \phi_R)^2,}
where \pl ~and \pr ~are left and right moving chiral bosons.  The system
has length $2R$ and its excitations have velocity $v$.  Henceforth
we set $v=1$.

We now allow the right and left movers to interact.
This interaction is realized through
an impurity at the origin:
\eqn\eIVii{
H_{imp} = -\left[ {\alpha \over 2} e^{i\pl - i\pr} +
{\alpha \over 2} e^{i\pr - i\pl} \right]_{x=0}.}
This impurity scatters left movers of charge $\nu e$
into right movers of the same charge.
For the value of $\nu$ we are interested in, it is the only relevant
operator which can induce scattering \kf .

Mapping $H_o + H_{imp}$ onto
boundary sine-Gordon is done in two steps \wa \fls .
First a change of basis is made and a spurious degree of freedom is
removed.  Secondly, the system is folded, changing it from an impurity
problem to a boundary problem.  As $H_{imp}$ depends only on the combination
$\pl -\pr$, the following change of basis is suggested:
\eqn\eIViii{
\phi^{e/o} = \left( \pl (x) \pm \pr (-x) \right).}
Both \pe ~and \po ~are left movers.  In this basis $H=H_0 +H_{imp}$ becomes
\eqn\eIViv{
H = -{1 \over 16\pi\nu} \int^R_{-R} dx (\partial_x \pe)^2 +
(\partial_x \po)^2 - \left[ {\alpha \over 2} e^{i\po (x=0)} +
{\alpha \over 2} e^{-i\po (x=0)} \right]}
We see that while \po ~is interacting \pe ~is not. \pe ~can thus be dropped
with H reducing to:
\eqn\eIViv{
H = -{1 \over 16\pi\nu} \int^R_{-R} dx (\partial_x \po)^2
- \left[ {\alpha \over 2} e^{i\po (x=0)} +
{\alpha \over 2} e^{-i\po (x=0)} \right]}
This is the first step of the mapping.

To implement the second step, folding the system, we separate out the degrees
of freedom of \po ~defined on $-R<x<0$ from \po ~defined on $0<x<R$ through
the variables:
\eqn\eIVv{\cases{
\pll (x) = \po (x) & $0 < x < R$ \cr
\plr (-x) = -\po (x) & $-R < x < 0$ .\cr}}
\pll ~(\plr) ~is a left (right) chiral boson.  The minus sign present
in the definition of \plr ~transforms the boson in the unfolded theory,
$\pl - \pr$, into its dual folded counterpart, $\pll + \plr$.
Under the change of variables, H becomes
\eqn\eIVvi{
H = -{1 \over 16\pi \nu}\int^R_0 dx (\partial_x \plr )^2 +
(\partial_x \pll )^2 - {\alpha \over 2}
\left[ e^{i\pll (x=0)} + e^{-i\pll (x=0)} \right]}
Because of the boundary, \pll ~and \plr ~are not independent.
In the limit $\alpha = 0$ we have \akl
\eqn\eIVvii
{\pll (x=0) = \plr (x=0) - \sigma .}
Setting $\sigma$ to zero and
treating the boundary term in conformal perturbation theory then allows us to
write
\eqn\eIVviii{
H = {1 \over 32\pi \nu}\int^R_0 dx (\partial_z \Phi )^2 +
(\partial_{\bar z} \Phi)^2 - \alpha \cos ( {\Phi \over 2} )\vert_{x=0},}
where $\Phi = \pll + \plr$.  For $\nu = 1/2$ this is precisely the
Hamiltonian
of boundary sine-Gordon at the free-fermion point.

To compute the ac-conductance of the system we use a Kubo type formula:
\eqn\eIVix{
G(w) = {1 \over (2R)^2} {i e^2 \over \hbar w} \int^R_{-R} dx dx'
\int^{\beta}_0 d\tau e^{iw_m\tau} \langle j_1 (x,\tau ) j_1 (x',0)
\rangle\vert_{w_m = -iw + \ep}}
where $\beta = 4\pi l$ is the inverse temperature and
$\langle j_1 (x,\tau ) j_1 (x',0)\rangle$ is a temperature Green function
of the spatial current in the unfolded system.  As such we need to
relate it to the correlators in the folded system.  $j_1$ is given by
\eqn\eIVx{
j_1 = -{i \over 2\pi} \partial_t \left( \pl - \pr \right).}
Hence in terms of the currents in the folded system,
$j_l = -{i\over 2\pi}\partial_t \pll$, $j_r = -{i\over 2\pi}\partial_t \plr$,
we have
\eqn\eIVxi{\eqalign{
\int^R_{-R} dx dx' \langle j_1 (x,\tau )  j_1 (x' ,0) \rangle =
\int^R_0 dx dx' & \langle j_l (x,\tau ) j_l (x', 0) \rangle
+ \langle j_r (x,\tau ) j_r (x', 0) \rangle \cr
& \langle j_l (x,\tau ) j_r (x', 0) \rangle +
\langle j_r (x,\tau ) j_l (x', 0) \rangle}}
The correlators on the r.h.s. of \eIVxi ~are obtained from those in section
3 by interchanging time and space (in this case the
boundary is in space).  With $x,x' > 0$, we obtain
\def\xk{(x' , 0)}
\eqn\eIVxii{\eqalign{
\langle j_r \xi j_r \xk \rangle &=
{1 \over (4l\pi )^2}
\left[\theta (x-x') {e^{-\bar w /2l} \over (1 - e^{-\bar w/2l})^2}
+ \theta (x'-x ) {e^{\bar w /2l} \over (1 - e^{\bar w/2l})^2} \right] ;\cr
\langle j_l \xi j_l \xj \rangle &= {1 \over (4l\pi )^2}
\left[\theta (x-x' ) {e^{-w/2l} \over (1 - e^{-w/2l})^2}
+ \theta (x'-x) {e^{w/2l} \over (1 - e^{w/2l})^2} \right] ;\cr
\langle j_r \xi j_l \xj \rangle &= {1 \over (4l\pi)^2}
\sum_{k,k'\in Z^+} {1 - \lambda_k\lambda_{k'} \over
(1 + \lambda_k)(1 + \lambda_k')}
e^{-(k+k'-1)\bar u/2l};\cr
\langle j_l \xi j_r \xj \rangle &=
{1 \over (4l\pi)^2}\sum_{k,k'\in Z^+} {1 - \lambda_k\lambda_{k'} \over
(1 + \lambda_k)(1 + \lambda_k')}
e^{-(k+k'-1) u/2l};\cr }}
where $w = x -x' + i\tau$ and $u = x+x' +i\tau$.

By taking the
Matsubara decomposition of these correlators and then analytically
continuing, $w_n \rightarrow -iw + \ep$, we obtain
\eqn\eIVxiii{\eqalign{
\langle j_r (x) j_r (x') \rangle (w) &=
-{i w \over 2\pi}  \theta (x'-x) e^{iw(x'-x)} ;\cr
\langle j_l (x) j_l (x') \rangle (w) &=
-{i w \over 2\pi} \theta (x-x') e^{iw(x-x')} ;\cr
\langle j_r (x) j_l (x') \rangle (w) &= 0; \cr
\langle j_l (x) j_r (x') \rangle (w) &= {i \over 4 \pi l}\int^\infty_{-\infty}
dk (1-f(k))(1-f(2lw-k)) \times \cr
& ~~~~~~~~~~~~~~~{k(k-2lw) - \delta^2 \over (i(k-2lw) + \delta)
(\delta - ik)}(e^{-4\pi wl} - 1) e^{iw(x+x')}; \cr}}
where $f(k) = (1+e^{2\pi k})^{-1}$ and $\delta = \alpha^2 l$.
This last two expressions are derived in Appendix
B.  $\langle j_r j_l \rangle$ only has negative
Matsubara frequencies.  Hence in the analytic continuation
to the retarded Green function, it vanishes.

Putting everything together we find for $G(w)$:
\eqn\eIVix{\eqalign{
{\rm Re}G(w) &= {e^2 \over h} {\sin^2 (wR/2) \over R^2 w^2}
\left[ 1 + {e^{4\pi l w} - 1 \over 2lw} \times \right. \cr
&\left. ~~~\int^{\infty}_{-\infty}dk
{\left( (k^2 - 2lwk)^2 - \delta^4 \right) \cos (wR) -
2lw\delta \sin (wR) \over (k^2 + \delta^2)((k-2lw)^2+\delta^2)}
f(k)f(2lw-k) \right];\cr
{\rm Im}G(w) &= {e^2 \over 2 h} {1 \over R^2 w^2}
\Biggl[ Rw - \sin (wR) + (e^{4\pi l w} - 1)\sin^2 (wR) \times  \cr
& ~~~ \left. \int^{\infty}_{-\infty}dk
{2lw\delta \cos (wR) + ((k^2 - 2lwk)^2 - \delta^4)\sin (wR) \over
(k^2 + \delta^2)((k-2lw)^2+\delta^2)}
f(k)f(2lw - k) \right] ,\cr}}
where $\delta = \alpha^2 l$.

Though these expressions are well behaved
over the entire ranges
of $R$, $w$, and $T = (4\pi l)^{-1}$, the physics they embody is not
similarly valid.  When the system size $2R$ is smaller than a thermal
coherence length $\hbar v / k_B T$ ($4\pi l$ in our units), we expect the
physics of the three dimensional leads to which the sample is attached
to begin playing a role \kf.  Moreover the limit in which
the Kubo formula for $G$ was derived presumes $R w < 1$.  So taking
$wR$ small, and hence $wl$ small, the above expressions at lowest
order reduce to
\eqn\eIVx{\eqalign{
{\rm Re}G(w) &= {e^2 \over 2 h} \left[
\int^{\infty}_{-\infty} dk {k^2 \over k^2 + X^4}
{e^k \over (1+e^k)^2} + \pi wl \int^{\infty}_{-\infty}
{k^2 - X^4 \over k^2 + X^4} e^k {e^{k} - 1 \over (1+e^k)^3} \right], \cr
{\rm Im}G(w) &= {e^2 wR \over 12 h} , \cr}}
where $X^2 = 2\pi\delta$.  The dc limit of this agrees
with the calculation of the d-c conductance in \kf.

\newsec{Acknowledgements}
The author would like to thank A. LeClair for helpful discussions.

\vfill\eject
\appendix A {Calculation of the Partition Function in the Limit $R \gg 2l$}
\def\S{S_\pm}
\def\lep{\log (1\pm e^{4\pi l k})}
\def\lem{\log (1\pm e^{-4\pi l k})}

We are interested in calculating the sum
\eqn\ai{
S_\pm = 2\pi l \sum_k k + \sum_k \log (1 \pm e^{-4\pi l k} ), }
where the sum is over $k$'s satisfying
\eqn\aii{
1 + X(k) = 1 + e^{-2iRk}{(\alpha^2 - ik)^2\over (\alpha^2 + ik)^2} = 0 .}
This sum may be recast as an integral
\eqn\aiii{
\S = {1 \over 2\pi i} \int_C dk {X'(k)  \over 1 + X(k)}
\left[ 2\pi l k + \lem \right] , }
where the contour C is pictured below in Figure A1.

\vskip .5in
\centerline{\psfig{figure=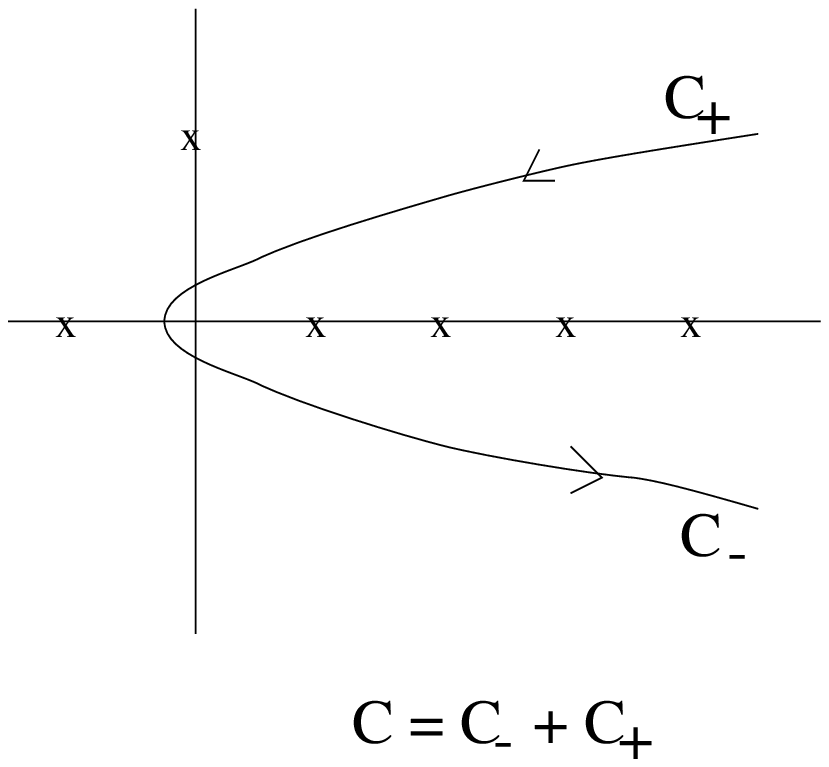,height=2in}}
\centerline{Figure A1.}
\vskip .5in

\noindent Because $X'(-k) = X'(k)/X^2(k)$ and
$X(-k) = X(k)^{-1}$, we can write
\eqn\aiv{\eqalign{
\int_{C_+} dk {X' \over 1+X}  & \left[2\pi l k +\lem ) \right] =
\int_{C_+} dk {X' \over X} \left[2\pi l k + \lem \right] + \cr
&\int_{-C_+} dk {X' \over 1 + X} \left[-2\pi l k + \lep \right] , \cr}}
where the contour $-C_+$ is pictured in Figure A2.

\vskip .5in
\centerline{\psfig{figure=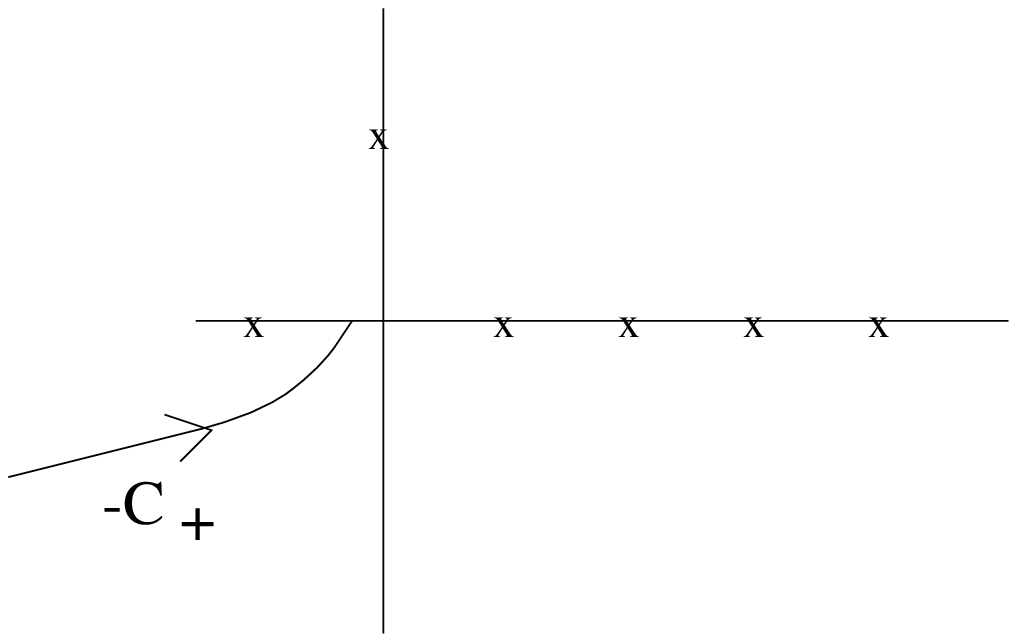,height=2in}}
\centerline{Figure A2.}
\vskip .5in

The second integral on the r.h.s. of the equation can be rewritten as
\eqn\av{\eqalign{
\int_{-C_+} dk {X \over 1 + X} & \left[-2\pi l k + \lep \right] = \cr
& \int_{-C_+} dk {X \over 1 + X} \left[2\pi l k + \lem \right] +
{(1\pm 1) \over 2}\pi i \log (2) .\cr}}
The sum now reduces to
\eqn\avi{\eqalign{
\S &= {1 \over 2\pi i} \int_{C_-} dk {X' \over 1+X}
\left[2\pi l k + \lem \right] \cr
& + {1 \over 2\pi i} \int_{C_+} dk {X' \over X}
\left[2\pi l k + \lem \right] \cr
& + {1 \over 2\pi i} \int_{-C_+} dk {X' \over 1+X}
\left[2\pi l k + \lep \right] + {(1\pm 1) \over 4}\log (2) .\cr}}
Combining the first and third integrals we obtain
\eqn\avii{\eqalign{
{1 \over 2\pi i} \int_{C_-} dk {X' \over 1+X}
&\left[2\pi l k + \lem \right]
 + {1 \over 2\pi i} \int_{-C_+} dk {X' \over 1+X}
\left[2\pi l k + \lem \right] \cr
& = {1 \over 2\pi i} \int_{C_- + (-C_+)} dk {X' \over 1+X} \lem \cr
& = {1 \over 2\pi i} \int_{C_o} dk
{\pm 4\pi l e^{-4\pi lk} \over 1\pm e^{-4\pi lk}} \log (1+X) \cr
& = \cases{\sum_{k=1/2,3/2,\cdots} \log (1+X(-ik/2l)),&\cr
\sum_{k=1,2,\cdots} \log (1+X(-ik/2l)),&,\cr}}}

\vskip .5in
\centerline{\psfig{figure=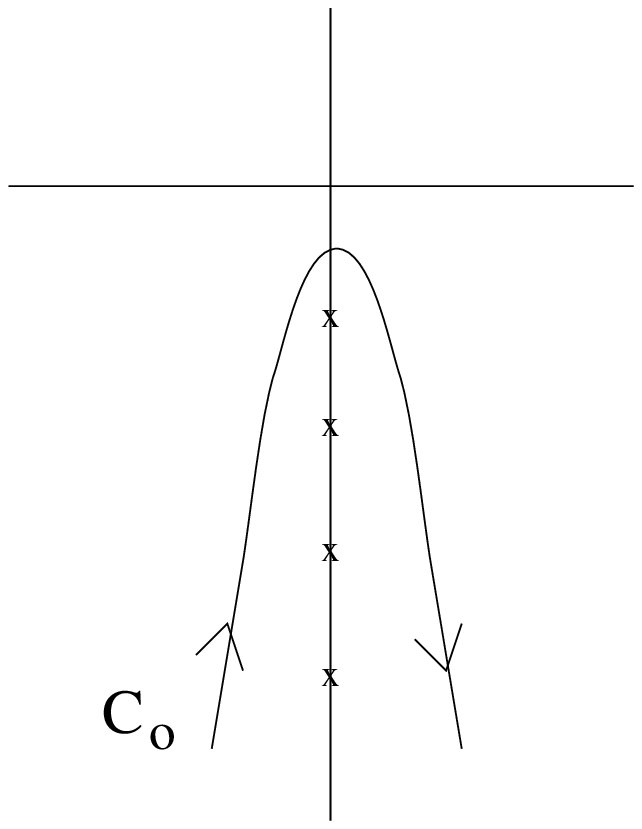,height=2in}}
\centerline{Figure A3.}
\vskip .5in

\noindent where in the second to last line we have integrated by parts and
deformed the contour to $C_o$ as defined in
Figure A3.  As $R \rightarrow \infty$ this last integral goes to zero.
It thus does not contribute to the boundary entropy.  So
$\S$ reduces to
\eqn\aviii{\eqalign{
\lim_{R \rightarrow \infty} S_\pm = -{1 \over 2\pi} \int_{C_+} dk
\left[ 2R + {4\alpha^2 \over \alpha^4 + k^2} \right]
&\left[ 2\pi lk + \lem \right] \cr
& + {1 \mp 1 \over 4} \log (2) .\cr}}
The terms proportional
to $2\pi lk$ arise from the ground state energy $-1/2\sum k$, and so
do not contribute to $g$.  Nor do the terms
proportional to $R$ contribute (g is solely a function of
boundary length).  So the contribution to g from $\S$ is
\eqn\avix{
{1 \over \pi} \int^\infty_0 dk { 1 \over 1 + k^2} \log (1\pm e^{-2\pi a k})
+  {1 \mp 1 \over 8} \log (2) ,}
where $a = 2l\alpha^2$.
\vfill\eject

\appendix B {Calculation of the $j_l-j_r$ Correlators}

We wish to evaluate
\eqn\bi{
\langle j_{r/l}(x) j_{l/r} (x') \rangle (w) =
{1 \over 16\pi^2 l^2} \left[ \int^\beta_0 d\tau e^{iw_n\tau}
I_\pm \right]_{w_n = -iw + \ep}}
where
\eqn\bii{\eqalign{
I_\pm &= \sum_{k,k' \in Z^+-1/2} { kk' - \delta^2 \over
(k+\delta )(k'+\delta )}e^{-(k+k')(x+x')/2l}e^{\pm i\tau (k+k'-1)/2l} ;\cr
\beta &= 4\pi l ;\cr
\delta &= \alpha^2 l . \cr}}
The sums $I_\pm$ can be written as contour integrals
\eqn\ebiii{
I_\pm = - \int_C dk \int_C dk' F_\pm (k,k') g(k,k') e^{\pm(k+k')\tau/2l}}
where
\eqn\biv{\eqalign{
F_\pm(k,k') &= (1-f(\mp k))(1-f(\mp k')) ;\cr
f(k) &= (1 + e^{2\pi k})^{-1} ; \cr
g(k,k') &= {-kk' - \delta^2 \over (-ik + \delta)(-ik' + \delta)}
e^{i(k+k')(x+x')/2l};}}
and the contour C is given below in Figure B1.
\vskip .5in
\centerline{\psfig{figure=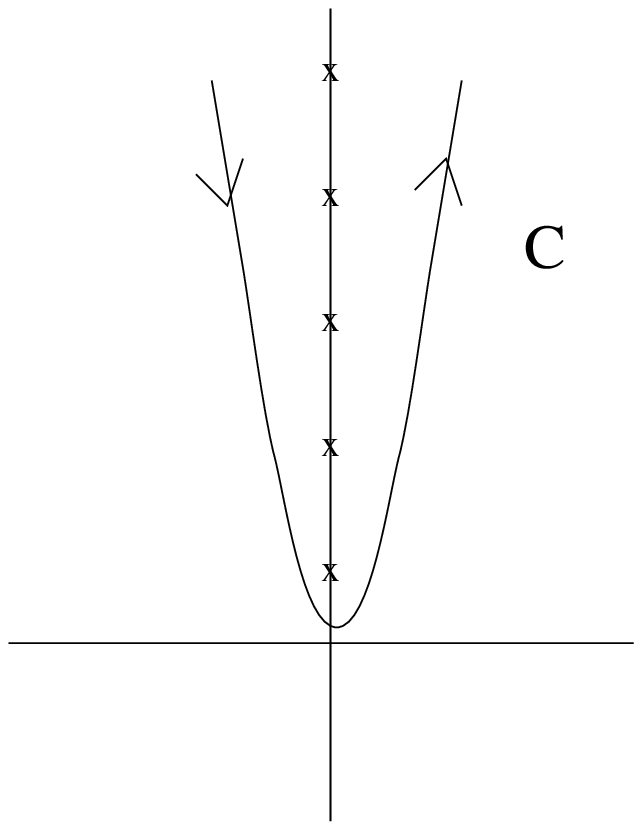,height=2in}}
\centerline{Figure B1.}
\vskip .5in
\noindent The Matsubara decomposition of $I_\pm$ equals
\eqn\bv{
I_\pm (w_n) = -2l
\int_C dk dk' F_\pm(k,k') g(k,k') {e^{\pm(k+k')2\pi} - 1 \over
i2lw_n \pm (k+k') },}
where $w_n = n/2l$, $n\in Z$.
Because we are going to make
the analytic continuation $w_n \rightarrow -iw + \ep$ we can assume
$w_n > 0$.  In this case $I_+$ is identically zero and $I_-$ may be
rewritten as
\eqn\bvi{\eqalign{
I_- (w_n>0)  = -2l
\int_C dk dk' F_\pm(k,k') & g(k,k') (e^{\pm(k+k')2\pi} - 1)
\times ~~~~~~~~~\cr
& \left[ {1 \over i2lw_n + (k+k')}  + {1 \over i2lw_n - (k+k')} \right]. }}
With $I_-$ in this form, the contours C can be continued to the real
axis and the analytic continuation made:
\eqn\bvii{\eqalign{
I_- (w) = - 2l
\int^\infty_{-\infty} dk dk' F_\pm(k,k') & g(k,k') (e^{\pm(k+k')2\pi} - 1)
\times ~~~~~~~~~ \cr
& \left[ {1 \over 2lw + i\ep + (k+k')}  + {1 \over 2lw + i\ep  - (k+k')}
\right]. }}
Having made the analytic continuation, we deform the contours back to C,
taking into account the pole at $k = w + i\ep - k'$:
\eqn\bviii{\eqalign{
I_- (w) = & - 2l
\int_C dk dk' F_\pm(k,k') g(k,k') (e^{\pm(k+k')2\pi} - 1)
 {4lw \over 4l^2w^2 - (k+k')^2} + \cr
& 4\pi l i (e^{-4\pi wl} - 1) \int^\infty_{-\infty} dk F_-(k,2lw-k)
g(k,2lw - k) .}}
The integral $\int_C$ vanishes identically because of the presence of
$(e^{-2\pi(k+k')} -1)$.  Hence we are left with only the second term.
The current-current correlators then reduce to the form claimed:
\eqn\bvix{\eqalign{
\langle j_r (x) j_l (x') \rangle (w) &= 0 ;\cr
\langle j_l (x) j_r (x') \rangle (w) &=
{i \over 4 \pi l} (e^{-4\pi wl} - 1)
\int^\infty_{-\infty} dk F_-(k,2lw-k) g(k,2lw-k).\cr}}

\listrefs

\bye